# PRECISE DETERMINATION OF THE WEAK MIXING ANGLE FROM A MEASUREMENT OF $A_{LR}$ IN $e^+e^- \to Z^0$

**M. Woods**

Stanford Linear Accelerator Center
Stanford University, Stanford, CA 94309

Representing

**The SLD Collaboration**†

**Abstract**

In the 1993 SLC/SLD run, the SLD recorded 50,000 $Z$ events produced by the collision of longitudinally polarized electrons on unpolarized positrons at a center-of-mass energy of 91.26 GeV. The luminosity-weighted average polarization of the SLC electron beam was (63.0±1.1)%. We measure the left-right cross-section asymmetry in $Z$ boson production, $A_{LR}$, to be 0.1628±0.0071(stat.)±0.0028(syst.) which determines the effective weak mixing angle to be $\sin^2\theta_W^{\rm eff} = 0.2292 \pm 0.0009(\text{stat.}) \pm 0.0004(\text{syst.})$.[1]

∗ Work supported in part by the Department of Energy, contract DE-AC03-76SF00515

† List of authors follows References.

## 1. Introduction

The left-right asymmetry is defined as $A^\circ_{LR} \equiv (\sigma_L - \sigma_R) / (\sigma_L + \sigma_R)$, where $\sigma_L$ and $\sigma_R$ are the $e^+e^-$ production cross sections for $Z$ bosons at the $Z$ pole energy with left-handed and right-handed electrons, respectively. The Standard Model predicts that this quantity depends upon the vector ($v_e$) and axial-vector ($a_e$) couplings of the $Z$ boson to the electron current,

$$A^\circ_{LR} = \frac{2v_e a_e}{v_e^2 + a_e^2} = \frac{2\left[1 - 4\sin^2\theta_W^{\mathrm{eff}}\right]}{1 + \left[1 - 4\sin^2\theta_W^{\mathrm{eff}}\right]^2}, \tag{1}$$

where the effective electroweak mixing parameter is defined[2] as $\sin^2\theta_W^{\mathrm{eff}} \equiv (1 - v_e/a_e)/4$.

Using the SLD detector, we count the number ($N_L$, $N_R$) of hadronic and $\tau^+\tau^-$ decays of the $Z$ boson for each of the two longitudinal polarization states (L,R) of the electron beam. The electron beam polarization is measured precisely with a Compton polarimeter. From these measurements we determine the left-right asymmetry,

$$A_{LR}(\langle E_{cm}\rangle) = \frac{1}{\left\langle \mathcal{P}_e^{lum}\right\rangle} \cdot \frac{N_L - N_R}{N_L + N_R} \tag{2}$$

where $\langle E_{cm}\rangle$ is the mean luminosity-weighted collision energy, and $\left\langle \mathcal{P}_e^{lum}\right\rangle$ is the mean luminosity-weighted polarization. This measurement does not require knowledge of the absolute luminosity, detector acceptance, or detector efficiency.

## 2. Polarized SLC Operation

The operation of the SLC with a polarized electron beam is illustrated schematically in Figure 1. Polarized electrons are produced by photoemission from a GaAs cathode. The electron spin orientation is longitudinal at the source and remains longitudinal until the transport to the Damping Ring (DR). In the Linac-to-Ring (LTR) transport, the electron spin precesses by 450° to become transverse at the entrance to the LTR spin rotator solenoid. This solenoid then rotates the electron spin to be vertical in the DR to preserve the polarization during the 8ms storage time. The spin orientation is vertical upon extraction from the DR; it remains vertical during injection into the linac and during acceleration to 46 GeV down the linac. The spin transmission of this system is 0.99, with the small loss resulting from the beam energy in the DR being 1.19 GeV, slightly lower than the design energy of 1.21 GeV; this causes the spin precession in the LTR to be 442° rather than 450°, and the spin transmission is the sine of this angle.

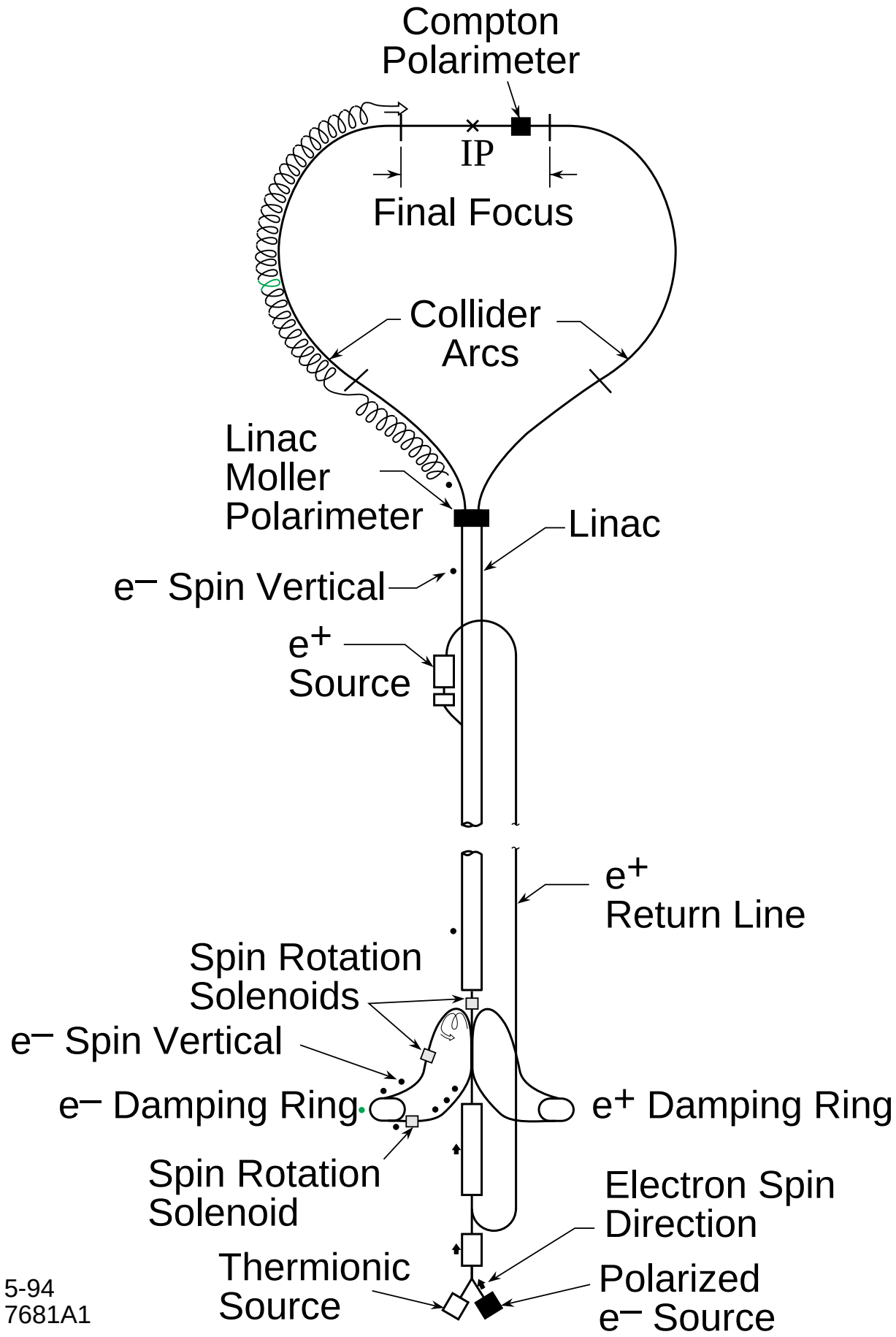

**Figure 1: The Polarized SLC**

The SLC Arc transports the electron beam from the linac to the SLC Interaction Point (IP) and is comprised of 23 achromats, each of which consists of 20 combined function magnets. The spin precession in each achromat is 1085°, while the betatron phase advance is 1080°. The SLC Arc is therefore operating near a spin tune resonance, with the effect that vertical betatron oscillations in the achromats, which rotate the beam direction about the vertical axis, can cause the beam polarization to rotate away from vertical; this rotation is a cumulative effect in successive achromats. (The rotation of the vertical spin component in a given achromat is simply due to the fact that rotations in $x$ and $y$ do not commute, while the cumulative effect is due to the spin resonance.) The resulting spin component in the plane of the arc then precesses significantly. We take advantage of the spin tune resonance and introduce a pair of vertical betatron oscillations ('spin bumps') to adjust the spin direction[3]. The amplitudes of these oscillations are empirically adjusted to achieve longitudinal polarization at the IP. Table 1 summarizes the relevant beam parameters for the 1993 run, and also gives the expected beam performance for the 1994 run.

**Table 1:** SLC Beam Parameters

| Parameter | 1993 | 1994 (expected) |
|---|---|---|
| $N^+$ | $3.0 \cdot 10^{10}$ | $(3.5\text{-}4.0)\cdot 10^{10}$ |
| $N^-$ | $3.0\cdot 10^{10}$ | $(3.5\text{-}4.0)\cdot 10^{10}$ |
| $f_{rep}$ | 120 Hz | 120 Hz |
| $\sigma_x$ | $0.8\mu m$ | $(0.4\text{-}0.5)\mu m$ |
| $\sigma_y$ | $2.6\mu m$ | $2.4\mu m$ |
| Luminosity | $5 \cdot 10^{29}$ cm$^{-2}$s$^{-1}$ | $(1-2) \cdot 10^{30}$ cm$^{-2}$s$^{-1}$ |
| Z/hr (peak) | 50 | 100-200 |
| Collision Energy | 91.26 GeV | 91.26 GeV |
| Polarization | 63% | 75% |
| Uptime | 70% | 70% |
| Run time | 6 months | 6 months |
| Integrated Zs | 50K | 100K - 150K |

## 3. Polarimetry at the SLC

The longitudinal electron beam polarization ($\mathcal{P}_e$) at the IP is measured by the Compton polarimeter[4] shown in Figure 1. This polarimeter detects Compton-scattered electrons from the collision of the longitudinally polarized electron beam with a circularly polarized photon beam; the photon beam is produced from a pulsed Nd:YAG laser operating at 532 nm. After the Compton Interaction Point (CIP), the electrons passes through a dipole spectrometer; a nine-channel Cherenkov detector then measures electrons in the range 17 to 30 GeV.

The counting rates in each Cherenkov channel are measured for parallel and anti-parallel combinations of the photon and electron beam helicities. The asymmetry formed from these rates is given by

$$A(E) = \frac{R(\rightarrow\rightarrow) - R(\rightarrow\leftarrow)}{R(\rightarrow\rightarrow) + R(\rightarrow\leftarrow)} = \mathcal{P}_e \mathcal{P}_\gamma A_C(E)$$

where $\mathcal{P}_\gamma$ is the circular polarization of the laser beam at the CIP, and $A_C(E)$ is the Compton asymmetry function. Measurements of $\mathcal{P}_\gamma$ are made before and after the CIP. By monitoring and correcting for small phase shifts in the laser transport line, we are able to achieve $\mathcal{P}_\gamma = (99 \pm 1)\%$. $A_C(E)$ and the unpolarized Compton cross-section are shown in Figure 2. The Compton spectrum is characterized by a kinematic edge at 17.4 GeV, corresponding to an 180° backscatter in the center of mass, and the zero-asymmetry point at 25.2 GeV. $A_C(E)$ is modified from the theoretical asymmetry function[5] by detector resolution effects. This effect is about 1% for the Cherenkov channel at the Compton edge. Detector position scans are used to locate precisely the Compton edge. The position of the zero-asymmetry point is then used to fit for the spectrometer dipole bend strength. Once the detector energy scale is calibrated, each Cherenkov channel provides an independent measurement of $\mathcal{P}_e$. The Compton edge is in channel 7, and we use this channel to determine precisely $\mathcal{P}_e$. The asymmetry spectrum observed in channels 1-6 is used as a cross-check; deviations of the measured asymmetry spectrum from the modeled one are reflected in the inter-channel consistency systematic error. Figure 3 shows the good agreement achieved between the measured and simulated Compton asymmetry spectrum.

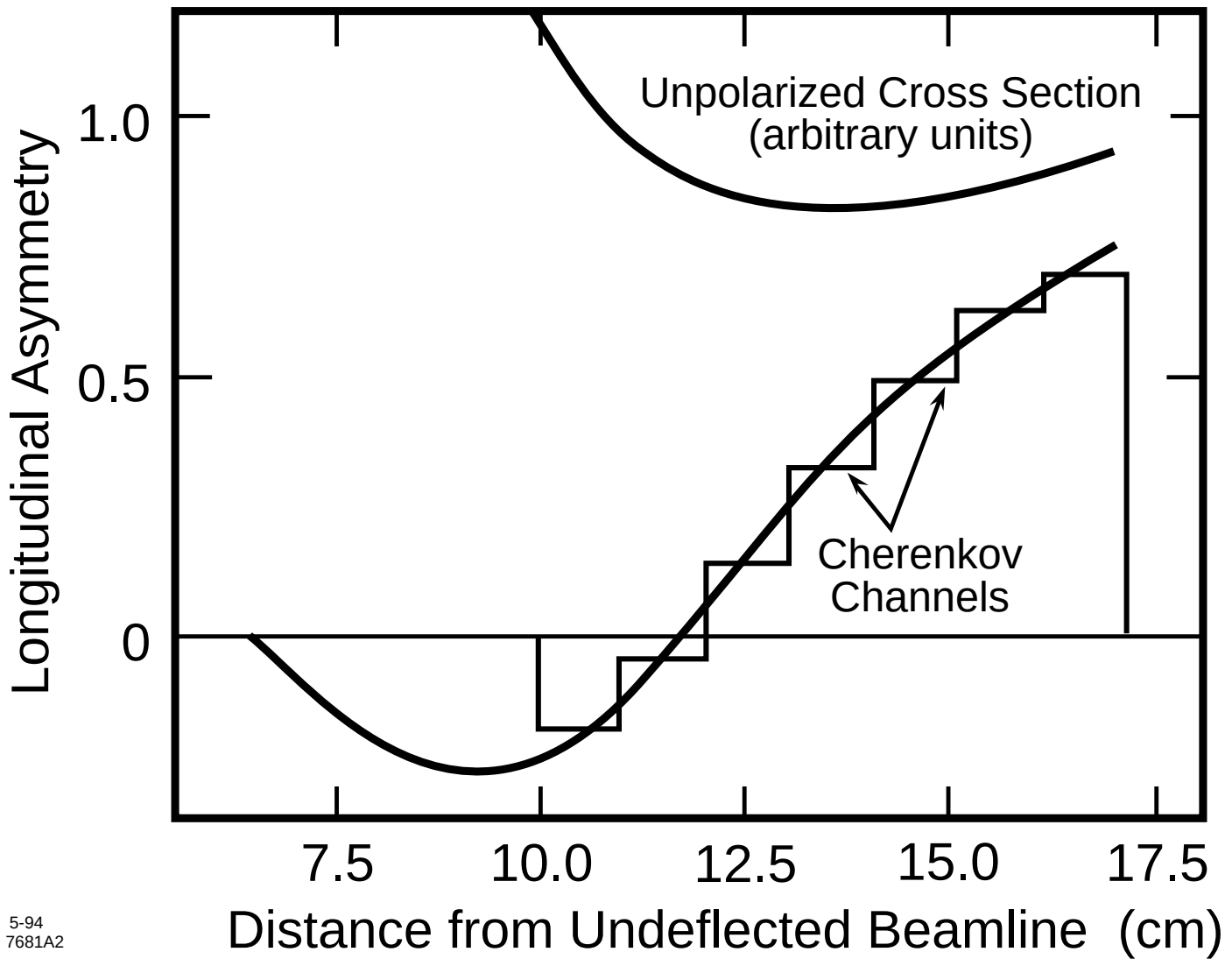

**Figure 2: Compton cross-section and asymmetry**

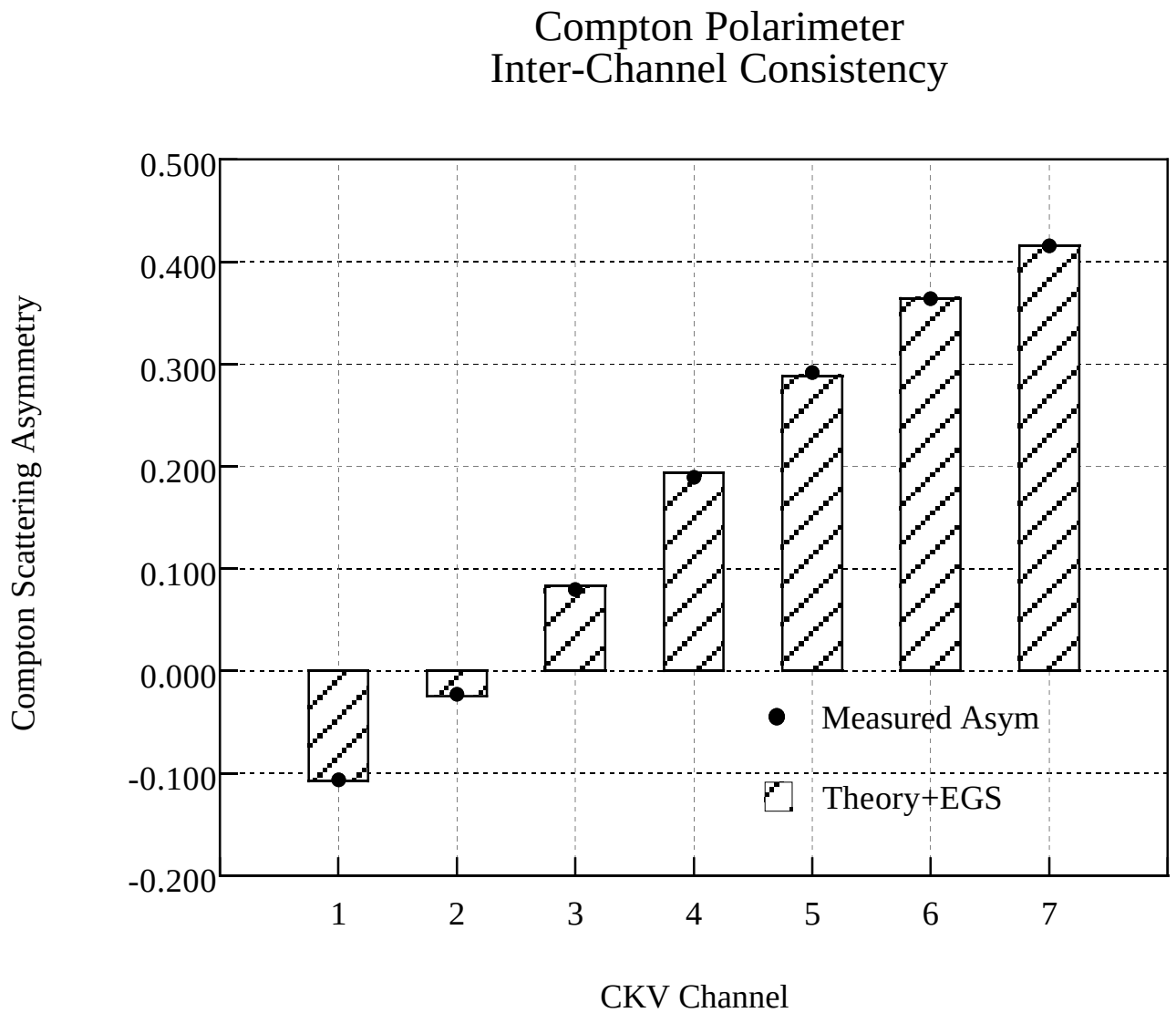

**Figure 3: Measured and simulated Compton asymmetry**

Polarimeter data are acquired continually during the operation of the SLC. The measured beam polarization is typically 61-64%. The absolute statistical precision attained in a 3 minute interval is typically $\delta\mathcal{P}_e = 1.0\%$. Averaged over the 1993 run, we find the mean beam polarization to be $\langle\mathcal{P}_e\rangle = (61.9 \pm 0.8)\%$. The systematic uncertainties that affect the polarization measurement are summarized in Table 2.

The Compton polarimeter measures $\mathcal{P}_e$ which can differ slightly from $\mathcal{P}_e^{lum}$. The main contribution to this difference arises from a chromatic effect. The electron beam is not monochromatic, but has an energy distribution, $N(E)$, which is characterized by a narrow core ($\Delta E/E <$ 0.2%) and a low-energy tail extending to $\Delta E/E \simeq -1\%$ defined by collimators at the end of the linac. The luminosity and beam polarization at the IP also have a dependence on energy given by $L(E)$ and $P(E)$. For the 1993 running, the energy dependence of $L(E)$ resulted from the small vertical spotsize and the resulting limitation on the luminosity due to third order chromatic aberrations in the final focus. $P(E)$ results from the effective number of spin rotations in the plane of the SLC Arc, which is measured to be 17.9 and is proportional to the energy. The three energy distributions $N(E), L(E), P(E)$ yield the beam polarization, $\mathcal{P}_e$ (ie. $P(E)$ weighted by by $N(E)$), and the luminosity-weighted beam polarization, $\mathcal{P}_e^{lum}$ (ie. $P(E)$ weighted by $N(E)\cdot L(E)$). These are related by $\mathcal{P}_e^{lum} \equiv \mathcal{P}_e(1+\xi)$, which defines the parameter $\xi$.

Off-energy electrons have reduced longitudinal polarization at the IP due to spin precession in the arc. They contribute less to the luminosity than on-energy electrons because they do not focus to a small spot at the IP; however they contribute the same as on-energy electrons to the Compton measurement of the beam polarization. Thus, $\mathcal{P}_e^{lum}$ can be greater than $\mathcal{P}_e$ due to this chromatic effect. $\mathcal{P}_e^{lum}$ is constrained, however, to be less than the polarization in the linac, $\mathcal{P}_e^{lin}$, since no spin precession occurs before the SLC Arc. Hence,

$$\mathcal{P}_e < \mathcal{P}_e^{lum} < \mathcal{P}_e^{lin} \qquad (3)$$

From measurements of $N(E)$ and $P(E)$ and from studies with a low-energy spread beam (core energy distribution rms less than 0.1% and no low-energy tail), we determine $(\mathcal{P}_e^{lin} - \mathcal{P}_e) <$ 4.7% relative. From measurements of $N(E)$ and $P(E)$ and from calculations of the worst-case distribution for $L(E)$, we determine $(\mathcal{P}_e^{lin} - \mathcal{P}_e^{lum}) > 1.4\%$ relative. Using equation (3), these two results constrain $\xi$ to be in the range from (0-3.3)%. Thus, we find $\xi = (1.7 \pm 1.1)\%$. We correct the Compton measurement of $\langle \mathcal{P}_e \rangle$ for this, and we find the luminosity-weighted polarization for the 1993 run to be $\langle \mathcal{P}_e^{lum} \rangle = (63.0 \pm 1.1)\%$.

The experiments described above to address the chromatic effect allow a determination of the beam polarization in the linac. We find this to be $\mathcal{P}_e^{lin} = (65.7 \pm 0.9)\%$ from measurements made on two separate occasions. This can be compared directly to measurements made by a diagnostic Moller polarimeter located at the end of the linac. This polarimeter analyzes the rate asymmetry in elastic scattering of the polarized electron beam from polarized electrons in a magnetized iron foil. After corrections for atomic momentum effects in the Moller target (the Levchuk effect) [6] we find the Moller measurements to give $\mathcal{P}_e^{lin} = (66 \pm 3)\%$, in good agreement with the Compton measurement.

Following the 1993 SLC/SLD run, the photocathode used for that run was taken to a test beamline with a newly commissioned Mott polarimeter. This polarimeter analyzes the rate asymmetry in elastic scattering of a polarized electron beam from nuclei in an uranium target. This polarimeter was built at UC Irvine and was calibrated there against another Mott polarimeter.[7] The SLAC Mott polarimeter measured the 1993 SLC photocathode to give a beam polarization of (64 ± 2)%, providing another cross-check on the Compton measurement.

## 4. $Z^\circ$ Event Selection

The $e^+e^-$ collisions are measured by the SLD detector which has been described elsewhere.[8] The trigger relies on a combination of calorimeter and tracking information, while the event selection is entirely based on the liquid argon calorimeter.[1] We estimate that the combined efficiency of the trigger and selection criteria is (93±1)% for hadronic $Z$ decays. Less than 1% of the sample consists of tau pairs. Because muon pair events deposit only small energy in the calorimeter, they are not included in the sample. The residual background in the sample is due primarily to beam-related backgrounds and to $e^+e^-$ final state events. We use our data and a Monte Carlo simulation to estimate the background fraction due to these sources to be (0.23 ± 0.10)%. The background fraction due to cosmic rays and two-photon processes is (0.02±0.01)%.

## 5. Measurement of $A_{LR}$

Applying the selection criteria, we find 27,225 ($N_L$) of the events were produced with the left-polarized electron beam and 22,167 ($N_R$) were produced with the right-polarized beam. The measured left-right cross section asymmetry for $Z$ production is

$$A_m \equiv (N_L - N_R)/(N_L + N_R) = 0.1024 \pm 0.0045.$$

To determine $A_{LR}$ from this, we use Equation (2) modified by some small correction terms,

$$A_{LR}(\langle E_{cm} \rangle) = \frac{A_m}{\langle \mathcal{P}_e^{lum} \rangle} + \frac{1}{\langle \mathcal{P}_e^{lum} \rangle} \left[ f_b(A_m - A_b) - A_{\mathcal{L}} + A_m^2 A_{\mathcal{P}} - E_{cm} \frac{\sigma'(E_{cm})}{\sigma(E_{cm})} A_E - A_\varepsilon \right] \quad (4)$$

where $f_b$ is the background fraction; $\sigma(E)$ is the unpolarized $Z$ cross section at energy $E$; $\sigma'(E)$ is the derivative of the cross section with respect to $E$; $A_b$, $A_{\mathcal{L}}$, $A_{\mathcal{P}}$, $A_E$, and $A_\varepsilon$ are the left-right asymmetries of the residual background, the integrated luminosity, the beam polarization, the center-of-mass energy, and the product of detector acceptance and efficiency, respectively.

The corrections defined in square brackets in equation (4) are found to be small. Of these corrections, the most significant one is that due to background contamination. The correction for this is moderated by a non-zero left-right background asymmetry ($A_b = 0.031 \pm 0.010$) arising from $e^+e^-$ final states which remain in the sample. Backgrounds give a net fractional correction to $A_{LR}$ of $(+0.17 \pm 0.07)\%$. Including all the corrections due to backgrounds and left-right asymmetries in luminosity, polarization, energy and efficiency gives a net correction to $A_{LR}$ of $(+0.10 \pm 0.08)\%$ of the uncorrected value.

Using equation (4), we find the left-right asymmetry to be

$$A_{LR}(91.26 \text{ GeV}) = 0.1628 \pm 0.0071(\text{stat.}) \pm 0.0028(\text{syst.}).$$

The contributions to the systematic error are summarized in Table 2. Correcting this result to account for photon exchange and for electroweak interference which arises from the deviation of the effective $e^+e^-$ center-of-mass energy from the $Z$-pole energy (including the effect of initial-state radiation), we find the effective weak mixing angle to be

$$\sin^2\theta_W^{\text{eff}} = 0.2292 \pm 0.0009(\text{stat.}) \pm 0.0004(\text{syst.}).$$

**Table 2:** Systematic uncertainties for the $A_{LR}$ measurement

| Systematic Uncertainty | $\delta\mathcal{P}_e/\mathcal{P}_e$ (%) | $\delta A_{LR}/A_{LR}$ (%) |
|---|---|---|
| Laser Polarization | 1.0 | |
| Detector Calibration | 0.4 | |
| Detector Linearity | 0.6 | |
| Interchannel Consistency | 0.5 | |
| Electronic Noise | 0.2 | |
| Total Polarimeter Uncertainty | 1.3 | 1.3 |
| Chromaticity Correction ($\xi$) | | 1.1 |
| Corrections in Equation (4) | | 0.1 |
| Total Systematic Uncertainty | | 1.7 |

## 6. Conclusions

We note that this is the most precise single determination of $\sin^2\theta_W^{\text{eff}}$ yet performed. Combining this value of $\sin^2\theta_W^{\text{eff}}$ with our previous measurement at $E_{CM} = 91.55$ GeV, we obtain the value, $\sin^2\theta_W^{\text{eff}} = 0.2294 \pm 0.0010$. This result can be compared to the determination of $\sin^2\theta_W^{\text{eff}}$ from measurements of unpolarized asymmetries at the $Z^0$ resonance performed by the LEP collaborations (Aleph, Delphi, L3, and OPAL). The LEP collaborations combine roughly 30 individual measurements of quark and lepton forward-backward asymmetries and of final state $\tau$-polarization, to give a LEP global average of $\sin^2\theta_W^{\text{eff}} = 0.2322 \pm 0.0005$.[9] The LEP and SLD results differ by 2.5 standard deviations.

The gauge structure of the Minimal Standard Model (MSM) of electro-weak interactions requires three parameters to specify it. These can be chosen to be $\alpha, G_F$ and $M_Z$.[10] The levels of precision being achieved at LEP and SLD for $\sin^2\theta_W^{\rm eff}$ and $\Gamma_Z$[10] are now providing significant tests of the MSM.[11] These tests will become even more stringent if the recent results for the top mass ($m_t = 174 \pm 16$ GeV) presented by the CDF collaboration[12] are confirmed, since the MSM predictions for $\sin^2\theta_W^{\rm eff}$ and $\Gamma_Z$ depend on $m_t$. In particular, the SLD $A_{LR}$ result considered within the MSM framework predicts the pole top mass to be $m_t = 250$ GeV $\pm$ 20 GeV (exp) $\pm$ 20 GeV ($m_H$), where the second error reflects the uncertainty for a Higgs mass in the range 60 GeV to 1000 GeV. This deviates from the CDF result by a little more than 2 standard deviations. Altarelli commented[13] in his presentation to this conference that the MSM 'works better than anticipated by many physicists.' Perhaps the level of precision in electroweak tests at LEP and SLC will soon be sufficient to reveal indications of physics beyond the MSM.

The next running period for SLD begins June 1, 1994 and the expected luminosity and beam polarization are given in Table 1. By the end of this run, the SLD expects to achieve a precision of 0.0005 for $\sin^2\theta_W^{\rm eff}$.

## REFERENCES

1. K. Abe et. al., SLAC-PUB-6456, March 1994. Submitted to *Phys. Rev. Lett.*
2. We follow the convention used by the LEP Collaborations in *Phys. Lett.* **B276**, 247 (1992).
3. T. Limberg, P. Emma, and R. Rossmanith, SLAC-PUB-6210, May 1993.
4. M.J. Fero *et al.*, SLAC-PUB-6423, April 1994.
5. See S.B. Gunst and L.A. Page, *Phys. Rev.* **92**, 970 (1953).
6. L.G. Levchuk, KHFTI-92-32, June 1992; and M. Swartz *et al.*, SLAC-PUB-6467, April 1994.
7. The calibration of the UC Irvine Mott polarimeter is described in H. Hopster, D.L. Abraham, *Rev. Sci. Instrum.* **59**, 49 (1988).
8. The SLD Design Report, SLAC Report 273, 1984.
9. The LEP results were presented by B. Pietrzyk at this conference. See also CERN-PPE/93-157, August 1993.
10. Precise results from the LEP collaborations on the mass and width of the $Z^0$ were presented at this conference by P. Clarke.
11. For a recent review see M. Swartz, SLAC-PUB-6384, November 1993; talk given at 16th Int. Symposium on Lepton and Photon Interactions, Ithaca, NY, August 1993. See also presentations to this conference by B. Pietrzyk, P. Clarke, F. Caravaglios, and G. Altarelli.
12. F. Abe et. al., FERMILAB-PUB-94-097-E, April 1994; submitted to *Phys. Rev. D.*
13. G. Altarelli, talk presented at this conference.

## †The SLD Collaboration

K. Abe,[27] I. Abt,[13] W.W. Ash,[25]† D. Aston,[25] N. Bacchetta,[20] K.G. Baird,[23]
C. Baltay,[31] H.R. Band,[30] M.B. Barakat,[31] G. Baranko,[9] O. Bardon,[16]
T. Barklow,[25] A.O. Bazarko,[10] R. Ben-David,[31] A.C. Benvenuti,[2] T. Bienz,[25]
G.M. Bilei,[21] D. Bisello,[20] G. Blaylock,[7] J.R. Bogart,[25] T. Bolton,[10]
G.R. Bower,[25] J.E. Brau,[19] M. Breidenbach,[25] W.M. Bugg,[26] D. Burke,[25]
T.H. Burnett,[29] P.N. Burrows,[16] W. Busza,[16] A. Calcaterra,[12] D.O. Caldwell,[6]
D. Calloway,[25] B. Camanzi,[11] M. Carpinelli,[22] R. Cassell,[25] R. Castaldi,[22][a]
A. Castro,[20] M. Cavalli-Sforza,[7] E. Church,[29] H.O. Cohn,[26] J.A. Coller,[3]
V. Cook,[29] R. Cotton,[4] R.F. Cowan,[16] D.G. Coyne,[7] A. D'Oliveira,[8]
C.J.S. Damerell,[24] S. Dasu,[25] F.J. Decker,[25] R. De Sangro,[12] P. De Simone,[12]
S. De Simone,[12] R. Dell'Orso,[22] Y.C. Du,[26] R. Dubois,[25] J.E. Duboscq,[6]
B.I. Eisenstein,[13] R. Elia,[25] P. Emma,[25] C. Fan,[9] M.J. Fero,[16] R. Frey,[19]
K. Furuno,[19] E.L. Garwin,[25] T. Gillman,[24] G. Gladding,[13] S. Gonzalez,[16]
G.D. Hallewell,[25] E.L. Hart,[26] Y. Hasegawa,[27] S. Hedges,[4] S.S. Hertzbach,[17]
M.D. Hildreth,[25] D.G. Hitlin,[5] J. Huber,[19] M.E. Huffer,[25] E.W. Hughes,[25]
H. Hwang,[19] Y. Iwasaki,[27] J.M. Izen,[13] P. Jacques,[23] J. Jaros,[25] A.S. Johnson,[3]
J.R. Johnson,[30] R.A. Johnson,[8] T. Junk,[25] R. Kajikawa,[18] M. Kalelkar,[23]
I. Karliner,[13] H. Kawahara,[25] M.H. Kelsey,[5] H.W. Kendall,[16] M.E. King,[25]
R. King,[25] R.R. Kofler,[17] N.M. Krishna,[9] R.S. Kroeger,[26] Y. Kwon,[25] J.F. Labs,[25]
M. Langston,[19] A. Lath,[16] J.A. Lauber,[9] D.W.G. Leith,[25] T. Limberg,[25]
X. Liu,[7] M. Loreti,[20] A. Lu,[6] H.L. Lynch,[25] J. Ma,[29] G. Mancinelli,[21]
S. Manly,[31] G. Mantovani,[21] T.W. Markiewicz,[25] T. Maruyama,[25] H. Masuda,[25]
E. Mazzucato,[11] J.F. McGowan,[13] A.K. McKemey,[4] B.T. Meadows,[8] R. Messner,[25]
P.M. Mockett,[29] K.C. Moffeit,[25] B. Mours,[25] G. Müller,[25] D. Muller,[25]
T. Nagamine,[25] U. Nauenberg,[9] H. Neal,[25] M. Nussbaum,[8] L.S. Osborne,[16]
R.S. Panvini,[28] H. Park,[19] T.J. Pavel,[25] I. Peruzzi,[12][b] L. Pescara,[20]
M. Piccolo,[12] L. Piemontese,[11] E. Pieroni,[22] K.T. Pitts,[19] R.J. Plano,[23]
R. Prepost,[30] C.Y. Prescott,[25] G.D. Punkar,[25] J. Quigley,[16] B.N. Ratcliff,[25]
T.W. Reeves,[28] P.E. Rensing,[25] L.S. Rochester,[25] J.E. Rothberg,[29] P.C. Rowson,[10]
J.J. Russell,[25] O.H. Saxton,[25] T. Schalk,[7] R.H. Schindler,[25] U. Schneekloth,[16]
D. Schultz,[25] B.A. Schumm,[15] A. Seiden,[7] S. Sen,[31] M.H. Shaevitz,[10]
J.T. Shank,[3] G. Shapiro,[15] D.J. Sherden,[25] C. Simopoulos,[25] S.R. Smith,[25]
J.A. Snyder,[31] M.D. Sokoloff,[8] P. Stamer,[23] H. Steiner,[15] R. Steiner,[1]
M.G. Strauss,[17] D. Su,[25] F. Suekane,[27] A. Sugiyama,[18] S. Suzuki,[18] M. Swartz,[25]
A. Szumilo,[29] T. Takahashi,[25] F.E. Taylor,[16] E. Torrence,[16] J.D. Turk,[31]
T. Usher,[25] J. Va'Vra,[25] C. Vannini,[22] E. Vella,[25] J.P. Venuti,[28] P.G. Verdini,[22]
S.R. Wagner,[25] A.P. Waite,[25] S.J. Watts,[4] A.W. Weidemann,[26] J.S. Whitaker,[3]
S.L. White,[26] F.J. Wickens,[24] D.A. Williams,[7] D.C. Williams,[16] S.H. Williams,[25]
S. Willocq,[31] R.J. Wilson,[3] W.J. Wisniewski,[5] M. Woods,[25] G.B. Word,[23]

J. Wyss,[20] R.K. Yamamoto,[16] J.M. Yamartino,[16] S.J. Yellin,[6] C.C. Young,[25] H. Yuta,[27] G. Zapalac,[30] R.W. Zdarko,[25] C. Zeitlin,[19] and J. Zhou,[19]

[1] *Adelphi University, Garden City, New York 11530*
[2] *INFN Sezione di Bologna, I-40126 Bologna, Italy*
[3] *Boston University, Boston, Massachusetts 02215*
[4] *Brunel University, Uxbridge, Middlesex UB8 3PH, United Kingdom*
[5] *California Institute of Technology, Pasadena, California 91125*
[6] *University of California at Santa Barbara, Santa Barbara, California 93106*
[7] *University of California at Santa Cruz, Santa Cruz, California 95064*
[8] *University of Cincinnati, Cincinnati, Ohio 45221*
[9] *University of Colorado, Boulder, Colorado 80309*
[10] *Columbia University, New York, New York 10027*
[11] *INFN Sezione di Ferrara and Università di Ferrara, I-44100 Ferrara, Italy*
[12] *INFN Lab. Nazionali di Frascati, I-00044 Frascati, Italy*
[13] *University of Illinois, Urbana, Illinois 61801*
[14] *KEK National Laboratory, Tsukuba-shi, Ibaraki-ken 305 Japan*
[15] *Lawrence Berkeley Laboratory, University of California, Berkeley, California 94720*
[16] *Massachusetts Institute of Technology, Cambridge, Massachusetts 02139*
[17] *University of Massachusetts, Amherst, Massachusetts 01003*
[18] *Nagoya University, Chikusa-ku, Nagoya 464 Japan*
[19] *University of Oregon, Eugene, Oregon 97403*
[20] *INFN Sezione di Padova and Università di Padova, I-35100 Padova, Italy*
[21] *INFN Sezione di Perugia and Università di Perugia, I-06100 Perugia, Italy*
[22] *INFN Sezione di Pisa and Università di Pisa, I-56100 Pisa, Italy*
[23] *Rutgers University, Piscataway, New Jersey 08855*
[24] *Rutherford Appleton Laboratory, Chilton, Didcot, Oxon OX11 0QX United Kingdom*
[25] *Stanford Linear Accelerator Center, Stanford University, Stanford, California 94309*
[26] *University of Tennessee, Knoxville, Tennessee 37996*
[27] *Tohoku University, Sendai 980 Japan*
[28] *Vanderbilt University, Nashville, Tennessee 37235*
[29] *University of Washington, Seattle, Washington 98195*
[30] *University of Wisconsin, Madison, Wisconsin 53706*
[31] *Yale University, New Haven, Connecticut 06511*
† *Deceased*
[a] *Also at the Università di Genova*
[b] *Also at the Università di Perugia*